\documentclass[journal]{IEEEtran}
\usepackage{textcomp}
\usepackage[utf8]{inputenc} 
\usepackage[T1]{fontenc}    
\usepackage{hyperref}       
\usepackage{url}            
\usepackage{booktabs}       
\usepackage{amsfonts}       
\usepackage{nicefrac}       
\usepackage{microtype}      
\usepackage{graphicx}
\usepackage{amsmath}
\usepackage{multirow,multicol}
\usepackage{color,colortbl} 
\usepackage{float}
\usepackage{eurosym}
\usepackage{amssymb}
\usepackage{array}
\usepackage{tabularx}
\usepackage{fancyhdr}
\newcommand{\markdown}[1]{#1}
\definecolor{Gray}{gray}{0.9} 
\def\BibTeX{{\rm B\kern-.05em{\sc i\kern-.025emb}\kern-.08em
    T\kern-.1667em\lower.7ex\hbox{E}\kern-.125emX}}
    
\begin{document}

\title{An Adaptive Intelligence Algorithm for Undersampled Knee MRI Reconstruction}

\author{%
  Nicola~Pezzotti$^\star$, Sahar~Yousefi$^\star$, Mohamed~S.~Elmahdy$^\star$, Jeroen~van~Gemert$^\star$, \\ Christophe~Sch\"ulke, Mariya~Doneva, Tim~Nielsen, Sergey~Kastryulin,\\  Boudewijn~P.F.~Lelieveldt, Matthias~J.P.~van Osch, Elwin de~Weerdt$^\dagger$, and Marius~Staring$^\dagger$
\thanks{$^\star$ denotes equal contribution.}
\thanks{$^\dagger$ denotes shared senior author.}
\thanks{N. Pezzotti,  C. Sch\"ulke,  M. Doneva, T. Nielsen, and S. Kastryulin are with Philips Research (e-mail: nicola.pezzotti@philips.com).}
\thanks{E. de~Weerdt and J.~van~Gemert are with Philips Healthcare, Best, The Netherlands.}
\thanks{S. Yousefi, M.S. Elmahdy, B. Lelieveldt, M.J.P. ~van Osch, and M. Staring are with the department of Radiology, Leiden University Medical Center, Leiden, The Netherlands.}}

\maketitle

\begin{abstract}
Adaptive intelligence aims at empowering machine learning techniques with the additional use of domain knowledge. In this work, we present the application of adaptive intelligence to accelerate MR acquisition. Starting from undersampled k-space data, an iterative learning-based reconstruction scheme inspired by compressed sensing theory is used to reconstruct the images. 
We developed a novel deep neural network to refine and correct prior reconstruction assumptions given the training data. The network was trained and tested on a knee MRI dataset from the 2019 fastMRI challenge organized by Facebook AI Research and NYU Langone Health. 
All submissions to the challenge were initially ranked based on similarity with a known groundtruth, after which the top 4 submissions were evaluated radiologically. Our method was evaluated by the fastMRI organizers on an independent challenge dataset. It ranked \#1, shared \#1, and \#3 on respectively the 8x accelerated multi-coil, the 4x multi-coil, and the 4x single-coil tracks. This demonstrates the superior performance and wide applicability of the method.
\end{abstract}


\section{Introduction}

\IEEEPARstart{M}{agnetic} Resonance Imaging (MRI) is a widely applied non-invasive imaging modality, with excellent soft tissue contrast and high spatial resolution. Unlike Computed Tomography (CT) scanning, MRI does not expose patients to any ionizing radiation, making it  a compelling alternative. 
MR images are essential for clinical assessment of soft tissue as well as functional and structural measurements, which leads to early detection and diagnosis of many diseases. However, MRI is relatively slow compared to other imaging modalities. The total examination time can vary from 15 minutes for knee imaging to an hour or more for cardiac imaging. Remaining still for this long in a confined space is challenging for any patient, being especially difficult for children, elderly and patients under pain. Motion artifacts are not only difficult to correct, which may require a complete re-scan~\cite{zaitsev2015motion}. Furthermore, the acquisition time affects the temporal resolution and subsequently limits the potential of MRI for dynamic imaging, where high temporal resolution and robustness against motion are critical for diagnosis. Moreover, the relatively long scan times lead to high costs that limit the availability of MRI scanners~\cite{cohen1991ultra}. Therefore, fast acquisition and reconstruction are crucial to improve the performance of  current MR scanners, which led in recent years to the development of techniques such as parallel reception, compressed sensing and multi-band accelerations. However, there is still a need for further scan acceleration.

The long acquisition time is intrinsic to the scanner and physics properties of MRI. For the majority of scans performed in clinical practice, this acquisition is done through consecutive reading-out of single lines in k-space. These readouts are constrained by physical limitations of the hardware, the contrast generating principle, and human physiology. 
The scanning time could be shortened by reducing the number of acquired lines in k-space, i.e. by undersampling the 2D or 3D k-space. However, this could violate the Nyquist criterion, resulting in aliasing and blurriness in the reconstructed images, rendering them unqualified for clinical purpose. Compressed Sensing (CS) and Parallel Imaging are the most common solutions for acceleration by undersampling, while maintaining image quality. Compressed Sensing, the focus of this paper, introduced by Donoho \cite{donoho2006compressed}, Lustig \cite{lustig2007sparse} and Candes~\cite{candes2011compressed}, leverages the fact that MR images can be compressed in some domain, restoring the missing k-space data through an iterative reconstruction algorithm~\cite{liang1992constrained}. Parallel Imaging uses multiple receive coils that provide an additional signal encoding mechanism, allowing to reduce the number of necessary k-space lines to reconstruct an image, thus partially parallelizing the data acquisition~\cite{pruessmann1999sense}.

When CS is used to accelerate MR acquisitions, the k-spaces is sampled pseudo-randomly  and the image is subsequently reconstructed by promoting a sparse solution. In the optimal setting, the reconstructed image will be identical to the Fourier transform of the full k-space and have a limited number of large coefficients when transformed to the sparse domain. Equation~(\ref{eq:basicL2L1}) shows the optimization function that describes the CS algorithm:
\begin{equation}
\min_{\mathbf{x}} \left\lbrace \parallel M \mathcal{F}\mathbf{x} - M\mathbf{y} \parallel ^{2} _{2} + \lambda \parallel \mathbf{\Psi}\mathbf{x} \parallel_{1} \right\rbrace,
\label{eq:basicL2L1}
\end{equation}
where $\mathbf{x}$ is the reconstructed image, $\mathbf{y}$ is the fully measured k-space data, $\mathcal{F}$ is the Fourier transform, $M$ (mask) is the undersampling operation, $\mathbf{\Psi}\mathbf{x}$ represents the sparsity transform coefficients, and $\lambda$ is the regularization parameter. The $\ell_1$ norm is used to enforce sparsity of the solution in a domain specified by the transformation $\mathbf{\Psi}$. The $\ell_2$ norm is used as a similarity measure between the measured k-space data $M \mathbf{y}$ and the reconstructed k-space $M \mathcal{F} \mathbf{x}$, called the ``data consistency'' term. Note that, in case of multi-coil acquisitions, the data consistency term is given by:
\begin{equation}
\sum_q \parallel M \mathcal{F}\left(\mathbf{S}_q \cdot \mathbf{x}\right) - M\mathbf{y}_q \parallel ^{2} _{2},
\label{eq:MultCoilL2}
\end{equation}
where $q$ denotes the coil element and $\mathbf{S_q}$ the corresponding coil sensitivity map. 
\markdown{The coil sensitivity maps $\mathbf{S}$ are computed using the fully centered region of k-space. A low-passed version of the coil images $\mathbf{x}^\text{lpf}_q$ is obtained by cropping the available region of k-space. The sensitivity map $\mathbf{S}_q$, for the individual coil element is computed as follows:
\begin{equation}
\mathbf{S}_q = \frac{ \mathbf{x}^\text{lpf}_q}{\sqrt{\sum_j \left(\mathbf{x}^\text{lpf}_j\right)^2}}
\label{eq:MultCoilL2}
\end{equation}
}
To simplify notation, without loss of generality, the single-coil data consistency term will be used throughout this paper.

Recently, deep learning has shown promising results for speeding up MR acquisition by adopting Convolutional Neural Networks (CNN) and Generative Adversarial Networks (GAN). In contrast to iteratively solving optimization problems, deep learning offers a solution for reconstructing highly-accelerated scans by adopting learnable reconstruction schemes. 

The literature of deep learning-based reconstruction algorithms can be divided into two categories~\cite{liang2019deep}. First, data-driven approaches, where a neural network is trained to find the optimal transformation from the zero-filled k-space to the desired reconstruction. Here, the network is completely dependent on the underlying training dataset without any task-specific prior knowledge on the domain; following are selected exemplar algorithms of this approach. Quan \emph{et al.} \cite{quan2018compressed} developed a GAN network for MR reconstruction starting from undersampled data. Their network consists of two consecutive networks, one for reconstruction and one for refining the results. They used a cyclic data consistency term alongside the WGAN loss. 
Mardani \emph{et al.} \cite{mardani2018deep} developed a GAN network for CS. The proposed network corrects aliasing artifacts of MR images. 
\markdown{Guo et al. \cite{guo2020deep} proposed a WGAN with recurrent context-awareness to reconstruct MRI images from highly
undersampled k-space data. Schlemper et al. propose a cascaded CNN-based compressive sensing (CS) technique for the reconstruction of diffusion tensor cardiac MRI~\cite{schlemper2018stochastic}.
Yang et al. proposed a conditional GAN-based architecture for de-aliasing and fast CS-MRI \cite{yang2017dagan, yu2017deep}.
}
Putzky \emph{et al.} \cite{putzky2019rim} treated the MR reconstruction problem as an inverse problem. They applied the previously introduced invertible Recurrent Inference Machine (i-RIM) model~\cite{putzky2019invert}, which iteratively updates its current state based on the output of the forward model. The model was trained and evaluated on the single- and multi-coil data at 4x and 8x accelerations from the fastMRI challenge (see Section \ref{sec:fastMRI_challenge} for more details).
AUTOMAP~\cite{zhu2018image} reports good reconstruction results with an architecture that learns to directly transform k-space into image data. Lee \emph{et al.} \cite{lee2018deep} introduced two separate deep residual networks for magnitude and phase. The proposed networks successfully reconstructed images even when obtained with high undersampling factors. 

Second, hybrid approaches are presented in the literature. This class of algorithms builds on top of existing reconstruction solutions and integrate learning-based approaches to substitute part of the original computations, often by adopting an unrolled implementation of an iterative algorithm~\cite{wang2018image}.
A notable example is the Variational network presented by Hammernik et al.~\cite{hammernik2018learning} utilizing learned filters in an existing iterative optimization scheme, while Yang et al. presented the Deep ADMM-Net~\cite{sun2016deep}, which extends the Alternating Direction Method of Multipliers (ADMM)~\cite{yang2010fast} approach by integrating learnable operators. 

Aggarwal \emph{et al.} \cite{aggarwal2018modl} introduced a model based deep learning architecture named MoDL to solve the inverse problem, including MR reconstruction. The proposed model consists of a series of recursive linear CNN networks. These networks share weights for regularization and reduction in the number of parameters. The proposed network imitates the CS algorithm and for numerical optimization, the authors introduced a data consistency term using a conjugate gradient (CG) optimization scheme at every iteration. The model was trained on multi-coil brain MR slices from 4 patients and tested on one patient. 
Ramzi \emph{et al.} \cite{ramzi2020benchmarking} provided a reproducible benchmark of deep learning based reconstruction methods on the single-coil part of the fastMRI dataset \cite{zbontar2018fastMRI}. The benchmark consists of a U-net \cite{ronneberger2015u}, cascade net \cite{schlemper2017deep}, KIKI-net \cite{eo2018kiki}, and PD-net \cite{adler2018learned}. Cascade net has been inspired by a dictionary learning approach \cite{caballero2014dictionary}. This approach is composed of residual convolutional blocks applied in image space followed by data consistency layers. The data consistency layers enforce the k-space values be close to the original k-space measurements. KIKI-net is \markdown{a cascaded network where a non-residual convolutional block has been added to perform k-space completion}, while PD-net provides a learnable and unrolled version of the Primal Dual Hybrid Gradient optimization algorithm \cite{chambolle2011first}.
\markdown{Seitzer et al. discussed the inadequacy of loss function for training a CS-MRI reconstruction CNN \cite{seitzer2018adversarial}. In that study they proposed a refinement method which incorporates both loss functions in a harmonious way to improve the training stability.}

Recently, Zhang and Ghanem \cite{zhang2018ista} developed a deep learning approach called ISTA-Net that mimics the conventional ISTA algorithm, but enriches it by replacing the sparsifying transform and the thresholding with learned operations.
The resulting network does not implement a fully iterative algorithm, but it simulates it by adopting a fixed number of iterations, effectively enabling the implementation of a deep neural network that can be trained by the backpropagation algorithm.
Inspired by the work of Zhang and Ghanem \cite{zhang2018ista}, in this paper we propose a deep-learning based solution, Adaptive-CS-Network, that mimics the ISTA algorithm, but introduces strong prior information, i.e., inductive biases, to better constrain the reconstruction problem. The main contributions of this work are: i) we propose a novel CNN network that integrates and enhances the conventional CS approach; ii) it integrates multiscale sparsification, inspired by wavelet transforms, but in a learnable manner; iii) we adopt domain-specific knowledge, such as data consistency, a prior on known phase behavior, and the location of the background: these computations cannot be easily learned by a CNN; iv) the proposed model exploits the correlation between neighbouring slices by adopting a 2.5D learning approach. In addition, we propose a hierarchical training strategy that leverages the available data. We conducted extensive experiments to investigate the performance of the network, and show that domain specific information is crucial for reconstructing high-quality MR images. The proposed network showed superior performance by winning one, and co-winning a second track out of the three tracks of the fastMRI challenge~\cite{zbontar2018fastMRI}.

\section{FastMRI Challenge}\label{sec:fastMRI_challenge}
The fastMRI challenge is a challenge organized by Facebook AI Research and NYU Langone Health \cite{zbontar2018fastMRI}. The aim of the challenge is to advance and encourage AI-based research in MR reconstruction in order to allow acceleration of the acquisition and, subsequently, to reduce the examination time. The challenge is divided in three tracks: 4x single-coil, 4x multi-coil, and 8x multi-coil accelerations. Eight teams participated in the multi-coil track and 17 teams in the single-coil track \cite{knoll2020advancing}. 

\subsection {Dataset}\label{sec:dataset}
The challenge organizers released a large-scale dataset of raw MR data of the knee \cite{knoll2020fastmri}. The data was acquired with a 2D protocol in the coronal direction with a 15 channel knee coil array using Siemens MR machines at two different field strengths: 1.5T and 3T \cite{zbontar2018fastMRI}. The data was acquired using two pulse sequences: a \emph{proton density weighting} with (PDFS) and without (PD) fat suppression. The data is divided approximately equally between these pulse sequences. The pixel size is 0.5 mm $\times$ 0.5 mm with a slice thickness of 3 mm. 

The dataset is divided in 4 categories: training (973 volumes, 34,742 slices), validation (199 volumes, 7,135 slices), test (118 volumes, 4,092 slices), and challenge (104 volumes, 3,810 slices). These numbers are the same for multi-coil and single-coil data, with the exception of the test and challenge categories, where single-coil data has respectively 10 and 12 volumes less than the multi-coil data. The training,  validation and test sets were publicly available since late November, 2018, while the challenge set was available since September 2019. The full k-space was available for all the datasets except for the test and challenge sets. Training and validation sets were considered for training and optimizing our model, while the test set was used for evaluating model performance on a public leaderboard. The final model was evaluated by the organizers on the independent challenge set.

The k-space data provided in the challenge were undersampled using a Cartesian mask, where k-space lines are set to zero in the phase encoding direction. The sampling density is dependent on the acceleration rate (4x or 8x), where the sampled lines are randomly selected. All masks, however, are fully sampled in the central area of k-space which corresponds to the low frequencies of the image. For the 4x accelerated scans, this percentage is 8\% while it is 4\% for 8x acceleration.
Besides making the reconstruction problem easier to solve, such lines allow for obtaining a low-pass filtered version of the image that is used to compute the coil \markdown{sensitivity} maps $\mathbf{S}_q$ as presented in Equation~(\ref{eq:MultCoilL2}) using a root sum of square approach~\cite{zbontar2018fastMRI}.

\subsection {Quantitative Evaluation}\label{ssec:quantitative}
In order to measure the accuracy of the reconstructed volumes $\mathbf{r}$ compared to the target volumes $\mathbf{t}$, the following metrics were considered: 

\subsubsection{Normalized Mean Square Error (NMSE)}measures the square of the Euclidean norm between a pair of images:
  \begin{equation}
    \mbox{NMSE} = \dfrac{\vert\vert \mathbf{r} - \mathbf{t}  \vert\vert^2_2}{\vert\vert \mathbf{t} \vert\vert^2_2}
    \label{eq:nmse}
    \end{equation}
  
\subsubsection{Peak Signal-to-Noise Ratio (PSNR)} the ratio between the maximum intensity and the underlying distortion noise:
\begin{equation}
\mbox{PSNR} = 10\log_{10} \dfrac{\max(\mathbf{t})^2}{ \frac{1}{N} \vert\vert \mathbf{r} - \mathbf{t}  \vert\vert^2_2}
  \label{eq:psnr}
\end{equation}

\subsubsection{Structural Similarity Index Metric (SSIM)}\label{sec:ssim}
measures image similarity using human perception aspects \cite{wang2004image}. SSIM is calculated by measuring three image distortions including luminance $l(\cdot)$, contrast $c(\cdot)$ and structure $s(\cdot)$:
    \begin{equation}
    \mbox{SSIM} = l(\mathbf{r},\mathbf{t})^\alpha c(\mathbf{r},\mathbf{t})^\beta s(\mathbf{r},\mathbf{t})^\gamma,
    \label{eq:ssim}
    \end{equation}
where $\alpha, \beta, \gamma$ are the distortion weights, here chosen as 1. \markdown{In this study, similar to the fastMRI challenge, the SSIM score is computed on the magnitude version of the 2D MR scans, leading to grayscale images.}
    
\subsection {Radiological Evaluation On The Challenge Dataset}
We submitted the reconstructions on the challenge dataset via an online form, which were then evaluated independently by the fastMRI organizers, described in detail by Knoll \emph{et al.}~\cite{knoll2020advancing}. All submissions were ranked by the SSIM metric, after which only the 4 highest ranking submissions were evaluated by a panel of 7 radiologists. The panel was asked to evaluate the reconstructions on a scale from 1 to 5 on four different categories, where 1 is the best and 5 is the worst. The 4 categories were the rating of artifacts, reconstruction sharpness, perceived contrast-to-noise ratio and diagnostic confidence. The radiological scores were subsequently averaged and translated to a final ranking.

\section{Methods}\label{sec:methods}
In this section we present the background of our solution, first by introducing the Iterative Shrinkage-Thresholding Algorithm (ISTA)~\cite{beck2009fast} and, second, by introducing its deep learning-based variant, ISTA-Net~\cite{zhang2018ista}.
Then, we present our solution, the Adaptive-CS-Network, that builds on top of the ISTA-Net framework by introducing several improvements, including strong inductive biases derived from domain knowledge on the reconstruction problem.

\subsection{ISTA Background}
ISTA is an optimization algorithm to solve \eqref{eq:basicL2L1} in an iterative fashion, starting from the reconstruction $\mathbf{x}_0$, which is often obtained by reconstructing the zero-filled undersampled k-space. The initial estimate is refined using the following update rules:
\begin{align}
\mathbf{r}_{i+1} &= \mathbf{x}_{i} - \rho \mathcal{F}^T (M \mathcal{F}\mathbf{x}_{i} - M\mathbf{y}),
\label{eq:data_consistency}\\
\mathbf{x}_{i+1} &= \arg \min_{\mathbf{x}} \frac{1}{2} \parallel  \mathbf{x} - \mathbf{r}_{i+1} \parallel ^{2} _{2} + \lambda \parallel \mathbf{\Psi}\mathbf{x} \parallel_{1},
\label{eq:ista_update}
\end{align}
where ${F}^T$ denotes inverse Fourier transform, $\mathbf{r}_{i+1}$ is an update of the estimate $\mathbf{x}_i$, where the error in the measured data $M\mathbf{y}$ is corrected by a step $\rho$.
Equation~\eqref{eq:ista_update} is a special case of the proximal mapping, with a regularization weight $\lambda$, and a crucial step for optimization algorithms such as ISTA, ADMM~\cite{yang2010fast} and AMP~\cite{donoho2009message}.
When $\mathbf{\Psi}$ is a wavelet transform $\mathbf{W}$, it can be proven that  
\begin{equation}
\mathbf{x}_{i+1} = \mathbf{W}^{-1} soft(\mathbf{W}\mathbf{r}_{i+1}, \lambda),
\label{eq:ista_soft_threshold}
\end{equation}
where $soft$ is the soft-tresholding operator defined as $soft(\mathbf{u},\lambda) = \max(|\mathbf{u}|-\lambda,0) \cdot \frac{\mathbf{u}}{|\mathbf{u}|}$.
In general, solving \eqref{eq:ista_update} is not straightforward for non-linear operators $\mathbf{\Psi}$, limiting the applicability of the ISTA framework to simple transforms. Another problem of this family of algorithms, is the difficulty of tuning the hyperparameters $\lambda$ and $\rho$ in addition to its slow convergence, hence requiring a lot of iterations to achieve the optimal solution of \eqref{eq:basicL2L1}.

\begin{figure*}
\begin{center}
\includegraphics[width=1\textwidth]{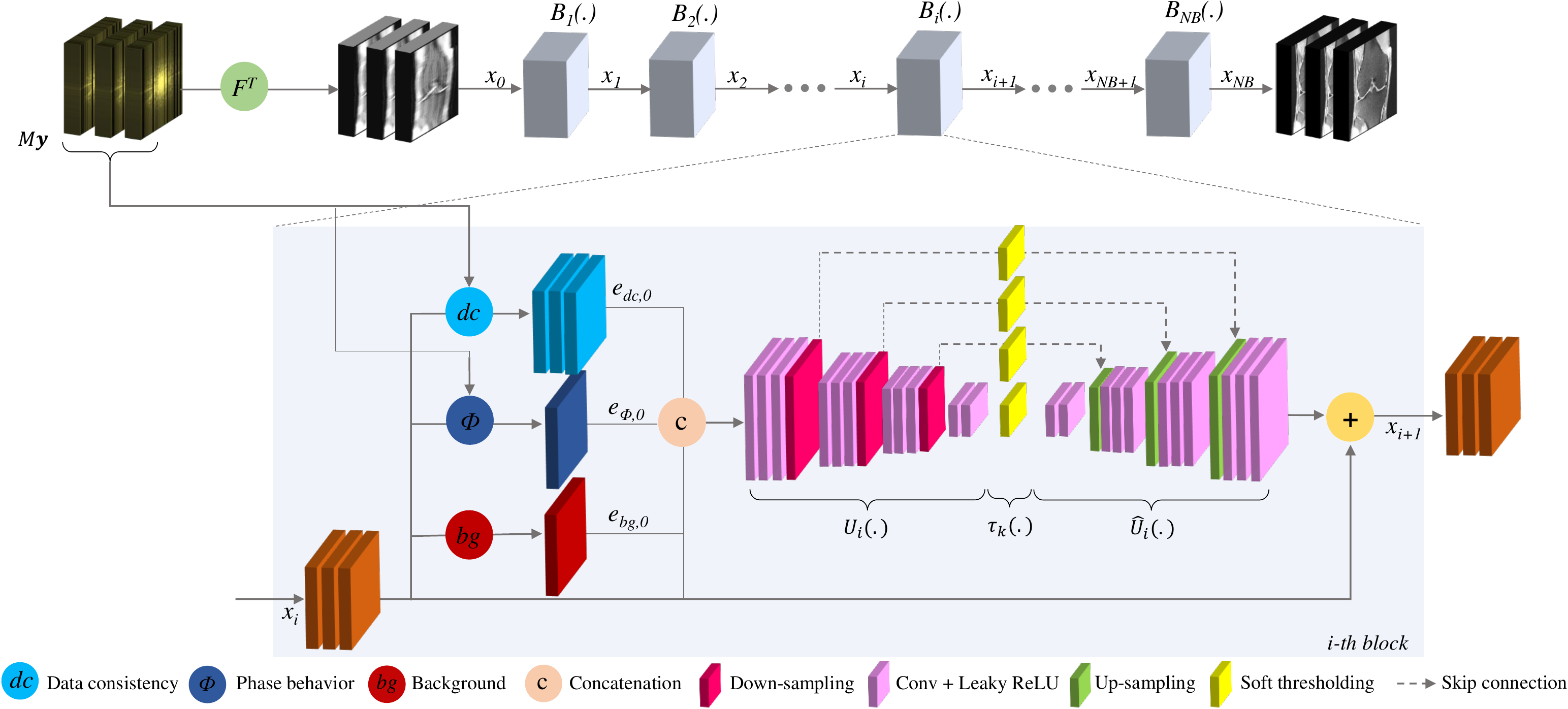}
\caption{Proposed adaptive Adaptive-CS-Net architecture. The input and output of the network are stacks of three consequent knee MR images.}
\label{fig:our_solution_1}
\end{center}
\end{figure*}

\subsection{ISTA-Net}
Recently, Zhang and Ghanem introduced a deep-learning approach to overcome the limitations of the ISTA framework for image-to-image reconstruction. 
Their solution, called ISTA-Net~\cite{zhang2018ista}, replaces the handcrafted transform $\mathbf{\Psi}$ with a learned operator $\mathcal{S}(\cdot)$, which consists of a 2D learnable convolution followed by a rectified linear unit (ReLU) and a second convolution.
By replacing $\mathbf{\Psi}$ with $\mathcal{S(\cdot)}$ in \eqref{eq:ista_update}, we can rewrite the update rule as
\begin{equation}
\mathbf{x}_{i+1} = \arg \min_{\mathbf{x}} \frac{1}{2} \parallel \mathbf{x} - \mathbf{r}_{i+1} \parallel ^{2} _{2} + \lambda \parallel \mathcal{S}(\mathbf{x}) \parallel_{1},
\label{eq:istanet_update}
\end{equation}
and, by defining $\hat{\mathcal{S}}$ as the inverse of $\mathcal{S}$, i.e., $\hat{\mathcal{S}} \circ \mathcal{S} = I$, Zhang and Ghanem propose to update \eqref{eq:ista_soft_threshold} as follows:
\begin{equation}
\mathbf{x}_{i+1} = \hat{\mathcal{S}}( soft(\mathcal{S}(\mathbf{r}_{i+1}), \lambda)),
\label{eq:ista_soft_threshold2}
\end{equation}
where $\hat{\mathcal{S}}$ has a similar architecture as $\mathcal{S}$.

The model is trained end-to-end, where the iterations of the ISTA algorithm are ``unrolled'', i.e., a number $b$ of identical reconstruction blocks are created. Note that in the ISTA-Net approach, the learnable parameters are shared among all the blocks in the unrolled network, unlike our solution.
The training loss is defined as a combination of the reconstruction and discrepancy loss:
\begin{align}
\label{eq:ista_loss}
\mathcal{L} &= \mathcal{L}_{reconstruction} + \sigma \mathcal{L}_{discrepancy} \\
\mathcal{L}_{reconstruction} &= \parallel \mathbf{x}_b - \mathcal{F}^T\mathbf{y} \parallel ^{2} _{2} \\
\mathcal{L}_{discrepancy} &= \frac{1}{b}\sum_{i=1}^{b}\parallel \hat{\mathcal{S}}( \mathcal{S}(\mathbf{x}_i)) - \mathbf{x}_i \parallel ^{2} _{2}
\end{align}
The reconstruction loss encodes the need for the final reconstruction, defined as $\mathbf{x}_b$, to be as close as possible in the least squares sense to the ground-truth image, i.e., $F^T\mathbf{y}$.
The discrepancy loss stimulates that $\hat{\mathcal{S}} \circ \mathcal{S} = I$. The $\sigma$ parameter allows to control the weight given to the discrepancy loss, and it is chosen to be arbitrarily small, e.g., $\sigma =0.01$.
An extension, called ISTA-Net$^+$ is also presented by the authors, where residual computations are adopted.

\subsection{Adaptive-CS-Network}
Starting from the network developed by Zhang and Ghanem, we developed the Adaptive-CS-Network approach. Our solution builds on top of the ISTA-Net solution based on three key innovations, here ordered by importance to the final network performance: i) the use of multi-scale and ii) multi-slice computations, together with iii) the introduction of soft MRI priors.
We present them independently, building towards the update rule of the Adaptive-CS-Network model as presented in \eqref{eq:acsnet_detailed}. Fig. \ref{fig:our_solution_1} illustrates the proposed network.

First, many non-learned CS algorithms make use of \textbf{multi-scale} transforms to sparsify the signal. 
An example is given in \eqref{eq:ista_soft_threshold}, where $\mathbf{W}$ is a wavelet transform; a decomposition of the signal into a set of basis functions at different scales.
We include this inductive bias in our design, and adopt a multi-scale transform $\mathcal{U}$, and its inverse $\hat{\mathcal{U}}$. As an additional design choice, we decide to sparsify and learn only the residual, therefore our update rule is written as follows:
\begin{equation}
\mathbf{x}_{i+1} = \hat{\mathcal{U}}( soft(\mathcal{U}(\mathbf{r}_{i+1}), \lambda_{s,f_s})) + \mathbf{r}_{i+1},
\label{eq:ista_soft_threshold3}
\end{equation}
where $\mathcal{U}$ comprises of 2D convolutions and non-linearities in the form of Leaky-ReLU to counteract the problem of dying neurons.
To generate a multiscale representation, a max-pooling layer is used and the resulting features are then processed again by convolutional blocks and non-linearities.
The exact design of $\mathcal{U}$ is presented in Fig.~\ref{fig:our_solution_1}.
The feature maps produced at the different scales are then thresholded using the soft-max function. 
Differently from ISTA-Net$^+$, we learn a lambda parameter and feature channel $f_s$ for each scale $s$. This approach gives the network the flexibility of tuning the thresholds independently, hence reducing the complexity of the transforms learned by the convolutional operators.
Finally, the filtered channels are transformed back into the image domain by the inverse $\hat{\mathcal{U}}$, consisting of interpolation, 2D convolutions and \markdown{Leaky-ReLU} operators.
Note that, contrary to the latest literature in deep learning networks, we decided not to adopt strided convolutions for sub- and up-sampling, which would increase the risk of creating checkerboard artifacts~\cite{araujo2019computing}; instead we took the more conservative approach of adopting pooling and interpolation layers for achieving better image quality.
Overall, the computation represented by $\hat{\mathcal{U}}( soft(\mathcal{U}(\mathbf{r}_{i+1}), \lambda_{s,f_s}))$ is implemented with a UNet-like architecture~\cite{ronneberger2015u}, where the feature maps before the skip connections are filtered according to the parameter $\lambda_{s,f_s}$.

Second, it is important to note that the slice thickness of the dataset is much higher than the in-plane resolution. This indicates that inter-slice correlations are less useful for finer scales, and potentially damaging as they will become a confounder for the network. However, such information becomes beneficial at coarser scales, e.g., to facilitate the delineation of the bone in several slices. Since our transform $\mathcal{U}$ is multi-scale by nature, we found it beneficial to inject \textbf{neighboring slices} into the model, while leaving it to the network to identify at which scale the information will be used. 
To reduce the memory footprint of the model, we adopted a 2.5D convolution approach by concatenating neighbouring slices into the input tensor along the channel dimension, enabling to ``reinvest'' the saved GPU memory as compared to a truly 3D convolution approach, into more unrolled iterations.
More details on the number of slices used and the definition of the loss function are given in Section~\ref{sec:multi_slice}.

Finally, we adopted a hybrid- or nudge- approach to incorporate additional \textbf{prior knowledge} into the reconstruction algorithm. We therefore computed additional information derived from the current estimate $\mathbf{x}_i$ together with k-space $M\mathbf{y}$. These soft priors, which are presented in the next section, capture some properties of an MR image that cannot be easily learned by a deep neural network due to the limited size of the receptive field.
The priors come in the form of images, and are provided as extra input channel to the transform $\mathcal{U}$.
In this way, they are integrated in the computations performed by $\mathcal{U}$ whenever this is beneficial for the optimization of the loss function.

\subsection{Final Design}

The overall update for a block $B_{i+1}$ in the Adaptive-CS-Network model is defined as follows:
\begin{align}
\begin{split}
\mathbf{x}_{i+1} &= B_{i+1}(\mathbf{x}_i) =  \\
  &\mathbf{x}_i + \hat{\mathcal{U}}_i \left(soft \left( \mathcal{U}_i \left( \mathbf{x}_i, \mathbf{e}_{\mathrm{dc},i}, \mathbf{e}_{\phi,i}, \mathbf{e}_{\mathrm{bg},i} \right), \lambda_{s,f_s}\right)\right).
\end{split}
\label{eq:acsnet_detailed}
\end{align}
Each block in the network learns different transforms $\mathcal{U}_i$ and $\hat{\mathcal{U}}_i$, enabling each block to focus on different properties of the image and effectively increasing the network capacity. Note that $\mathcal{U}_i$ and $\hat{\mathcal{U}}_i$ are different for every reconstruction block $i$.

In our final design, the transform $\mathcal{U}_i$ does not receive the data consistent image $\mathbf{r}_i$, as defined in \eqref{eq:data_consistency}, but rather the current estimate $\mathbf{x}_i$ together with the data consistency prior $\mathbf{e}_{\mathrm{dc},i}$ computed as follows:
\begin{equation}
\mathbf{e}_{\mathrm{dc},i} = \mathcal{F}^T (M \mathcal{F}\mathbf{x}_{i} - M\mathbf{y}).
\label{eq:priors_dataconsistency}
\end{equation}
This ``soft data-consistency'' update allows the network to evaluate the reliability of the acquired data and potentially compensate errors in the coil combination defined by $\mathcal{F}$ in \eqref{eq:basicL2L1}.

The second prior we provided to the network, $\mathbf{e}_{\phi,i}$, represents the known phase response for spin-echo MR sequences. Theoretically, spin-echo sequences have zero phase everywhere in the image. In practice, however, slowly varying phase will occur, i.e. nonzero phase only in the low frequencies, due to hardware and acquisition imperfections. Taking this into account, it is noted that the final reconstructed image should be a real valued image after removal of the slowly varying phase. This information is captured in the following prior:
\begin{equation}
\mathbf{e}_{\phi,i} = \left\{\mathbf{x}_{i} \cdot \frac{\mathbf{x}^{*}_{i,\mathrm{lpf}}}{\|\mathbf{x}_{i,\mathrm{lpf}}\|_2}\right\}_{imag},
\label{eq:priors_spin_behaviour}
\end{equation}
where $^{*}$ denotes the complex conjugate, and $\mathrm{lpf}$ refers to low pass filtering. The low pass filter is chosen such that it corresponds to the center part of k-space which is fully sampled. By doing so, the low pass filtered image $\mathbf{x}_{i,\mathrm{lpf}}$ can be derived beforehand only once, hence $\mathbf{x}_{i,\mathrm{lpf}}$ is replaced by $\mathbf{x}_{0,\mathrm{lpf}}$.

Finally, we adopt a simple approach to estimate the location in $\textbf{x}_i$ where the background is found, which is common in parallel imaging techniques. The following prior is applied:
\begin{equation}
\mathbf{e}_{\mathrm{bg},i} = \frac{\mathbf{x}_{i}}{\| \mathbf{x}_{i,\mathrm{lpf}}\|_2 }.
\label{eq:priors_background}
\end{equation}
This prior will penalize estimated signal content where $\| \mathbf{x}_{i,\mathrm{lpf}}\|$ is low, i.e., within the background. Again, $\mathbf{x}_{i,\mathrm{lpf}}$ is replaced by $\mathbf{x}_{0,\mathrm{lpf}}$. Because $\mathbf{x}_{0,\mathrm{lpf}}$ is based on the fully measured central part of k-space, the image is artefact free albeit at low spatial resolution, leading to a reliable background identification.

In Fig.~\ref{fig:our_solution_1} the design of the Adaptive-CS-Network is shown, including the multi-scale transforms, the multi-slice computation and the priors provided as input. Note how the spin-echo and background priors are computed only for the central slice, in order to save GPU memory.


\subsection{Network Training and Implementation Details}

We implemented our models in PyTorch~\cite{paszke2017automatic}. All the optimization experiments were performed on \markdown{an} NVIDIA V100 GPU with 16 GB RAM and the final network was trained on two NVIDIA V100 GPU with 16 GB RAM. In order to run as many experiments as possible given the challenge deadline, model optimization (see  Section~\ref{sec:model_optimization}) was done with a relatively small model ($\leq$ 10 blocks), which we trained for 20 epochs.
All the optimization networks were trained and validated on the highest acceleration rate of the challenge, i.e. 8x and for single-coil data, except for the number of the blocks which was performed for both 4x and 8x, and for the priors which are more relevant for the multi-coil data. Since the ground truth for the test set was not available, all the quantitative comparisons were only done on the validation set. 

For the challenge, we trained the final model using the training and validation datasets for 25 epochs and accelerations randomly selected from 2x to 10x. \markdown{The residual connections designed on a per-iteration basis, facilitates the learning and prevents the degradation of the error gradient throughout the architecture}. The model was subsequently fine-tuned on eight data sub-populations identified by the acceleration (4x and 8x), the protocol (PD and PDFS) and the scanner field strength (1.5T and 3T). Fine-tuning was then performed for 10 epochs on the sub-populations. This procedure was performed independently for the single- and multi-coil datasets, resulting in a total of 8 models.
All models were trained using an exponentially decaying learning rate of $10^{-4}$. 
\markdown{The final models have 33M trainable parameters each; for the single-coil data this leads to an inference time of approximately 327 ms, while it takes approximately 518 ms to compute the reconstruction of a multi-coil dataset on an NVIDIA V100 GPU}.

\section{Experiments and results: Model optimization}\label{sec:model_optimization}

In this section we present how we optimized the network configuration, on a smaller model with $10$ reconstruction blocks, using the quantitative measures reported in Section \ref{ssec:quantitative} for validation. We performed experiments on the number of the blocks, the loss functions, the influence of using adjacent slices, the optimizer, and the soft priors. A repeated measure one-way ANOVA test was performed on the SSIM values using a significance level of $p = 0.05$. P-values are only stated for the comparisons between the best method and the other methods. In all the experiments a learning rate of 0.0001 was used.

\subsection{Number of blocks }\label{sec:phase_no}
\begin{figure}[!tb]
    \centering
    \includegraphics[width=.5\textwidth]{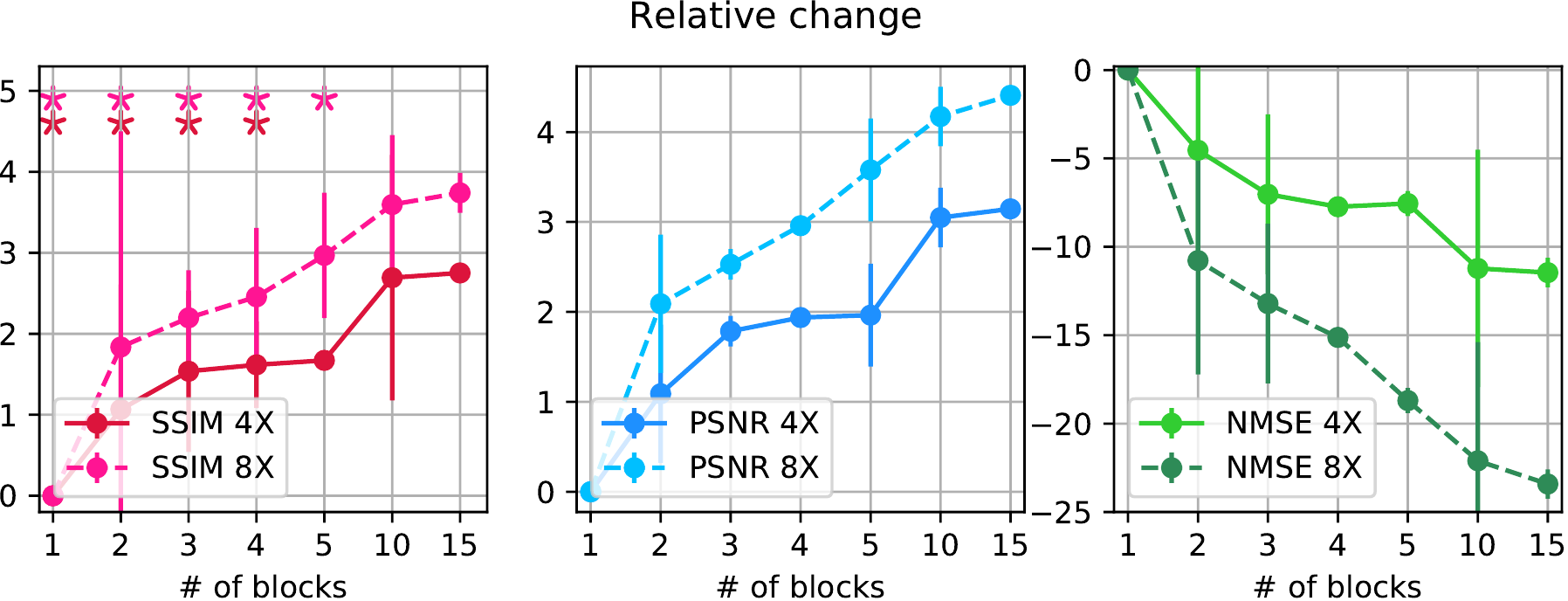}
    \caption{The effect of the number of blocks on performance, using the 4x and 8x single-coil validation data. The variance values are shown by the bars. The stars in the first plot show one-way ANOVA statistical significance.}
    \label{fig:phaseno}
\end{figure}

The proposed model consists of multiple blocks, related to the number of unrolled iterations of the ISTA scheme. Increasing the number of blocks leads to an increase in the number of parameters of the model, and subsequently training time and GPU memory usage as well as an increase in risk of overfitting. In this experiment we investigated the effect of the number of the blocks on the quality of reconstructed images. Tests were ran with the 2D network for 4x and 8x acceleration rates without neighboring slices, MSE as loss function, RMSprop as optimizer, and with the Unet-like architecture of 16 filter maps for each convolutional layer. 
Fig. \ref{fig:phaseno} reports the relative changes to a single block of our quantitative metrics. Based on the experiments, increasing the number of the blocks will improve the performance of the network. Therefore, the final network was configured with the maximum number of blocks that could be fitted into GPU memory: 25 blocks. However, for the optimization experiments below only 10 blocks were employed to limit the duration of the training.

\subsection{Loss functions}\label{sec:loss_func}

In this experiment we investigated the effect of a wide range of \markdown{differentiable} loss functions on the performance of our network.
Here, we used the single slice reconstruction network with only 10 blocks, RMSprop as the optimizer, and 16 filter maps for each convolutional layer. The models were trained for 20 epochs to ensure convergence of the model. The evaluated loss functions included MSE, perceptual loss (PL) \cite{johnson2016perceptual}, $\mathrm \ell_1$, Huber~\cite{huber1992robust}
and multi-scale structural similarity index (MSSIM)~\cite{wang2003multiscale}. The PL loss function was calculated using a pre-trained VGG-16 at layers relu1$\_$2, relu2$\_$2, and relu3$\_$3. 

\begin{table}[t]
\let\center\empty
\let\endcenter\relax
\centering
\caption{The effect of the loss function on performance, using the 8x single-coil validation data. Stars denote one-way ANOVA statistical significance.}\label{table:lossfuncs}
\resizebox{1.\width}{!}{
\begin{tabular}{lllllll}
 \hline
  \multirow{2}{*}{Loss function} & \multicolumn{1}{c}{SSIM} &  \multicolumn{1}{c}{NMSE} & \multicolumn{1}{c}{PSNR} & \multirow{2}{*}{$p$-value} \\
 & \multicolumn{1}{c}{$\mu\pm\sigma$} & \multicolumn{1}{c}{$\mu\pm\sigma$} & \multicolumn{1}{c}{$\mu\pm\sigma$} & \\
  \hline
\rowcolor{Gray} 
MSE & 0.657$\pm$0.149$^*$ & 0.046$\pm$0.029 & 30.2$\pm$2.8 &$\ll$0.001\\
Perceptual  loss & {0.664$\pm$0.157$^*$} & {0.061$\pm$0.044} & {29,2$\pm$3.2} &$\ll$0.001\\
\rowcolor{Gray}
Huber  & {0.664$\pm$0.148$^*$} & {0.062$\pm$0.041} & {29.1$\pm$3.0} &$\ll$0.001\\
$\mathrm \ell_1$ & {0.664$\pm$0.148$^*$} & {0.062$\pm$0.041} & {29.1$\pm$3.0} &$\ll$0.001\\
\rowcolor{Gray}
SSIM & {0.662$\pm$0.145$^*$} & {0.065$\pm$0.041} & {28.9$\pm$2.8} &$\ll$0.001\\
MSSIM \cite{wang2003multiscale} & {0.671$\pm$0.143$^*$} &  {0.050$\pm$0.034} & {30.1$\pm$3.1} &$\ll$0.001\\
\rowcolor{Gray}
Eq. (\ref{eq:loss_mssim_l1})  & \textbf{0.673$\pm$0.143} &  \textbf{0.048$\pm$0.033} & \textbf{30.3$\pm$3.1}& \\
\hline
\end{tabular}
}
\end{table}

MSSIM \cite{wang2003multiscale} builds upon SSIM (see Section \ref{sec:ssim}) by incorporating structural similarity at multiple image resolutions, thereby supplying more flexibility compared to SSIM, and is defined as follows:
\begin{equation}
    \text{MSSIM} = \left[l_M(\mathbf{r^c},\mathbf{t^c})\right]^{\alpha_M} \prod_{i=1}^{M} \left[c_i(\mathbf{r^c},\mathbf{t^c})\right]^{\beta_i} \left[s_i(\mathbf{r^c},\mathbf{t^c})\right]^{\gamma_i},
\end{equation}
where $\mathbf{r}^c, \mathbf{t}^c,$ denote the reconstructed and target images respectively, $M$ is the number of scales used, $l_M$, c$_i$ and s$_i$ are the luminance, contrast, and structure as defined in~\cite{wang2004image}, $\alpha_M$, $\beta_i$, and $\gamma_i$ are the weights of the distortion factors at different resolution levels. We adopted the same weights as reported in \cite{wang2003multiscale}.

Zhao et al. \cite{zhao2016loss} reported that a linear combination of SSIM and $\ell_1$ preserves the different properties of an image better than each separately: SSIM encourages the network to preserve structural and contrast information, while $\ell_1$ enforces sensitivity to uniform biases to preserve luminance \cite{zhao2015loss}. Since MSSIM reached higher metric values than SSIM (see Table \ref{table:lossfuncs}), we deployed a weighted summation of MSSIM \cite{wang2003multiscale} and $\ell_1$:
\begin{equation}\label{eq:loss_mssim_l1}
L = \alpha \text{MSSIM}( \mathbf{r^c} , \mathbf{t^c} ) + (1-\alpha) \Vert \mathbf{r^c} - \mathbf{t^c} \Vert_1,
\end{equation}
where $\alpha = 0.84$ was chosen, following Zhao et al.   \cite{zhao2016loss}. 
Note that, compared to the ISTA-Net approach, we found it beneficial not to adopt the discrepancy loss as presented in Eq.~\eqref{eq:ista_loss} for two reasons.
First, we empirically found that tuning the loss multiplier $\theta$ is not straightforward, leading to sub-optimal results in terms of the reconstruction loss. Secondly, computing the discrepancy loss is very demanding in terms of GPU memory, requiring to perform a second forward pass where only the thresholding operation is ignored. While feasible, it requires to make the model significantly smaller in terms of learnable parameters, hence reducing model performance significantly. 

\begin{figure*}[!tb]
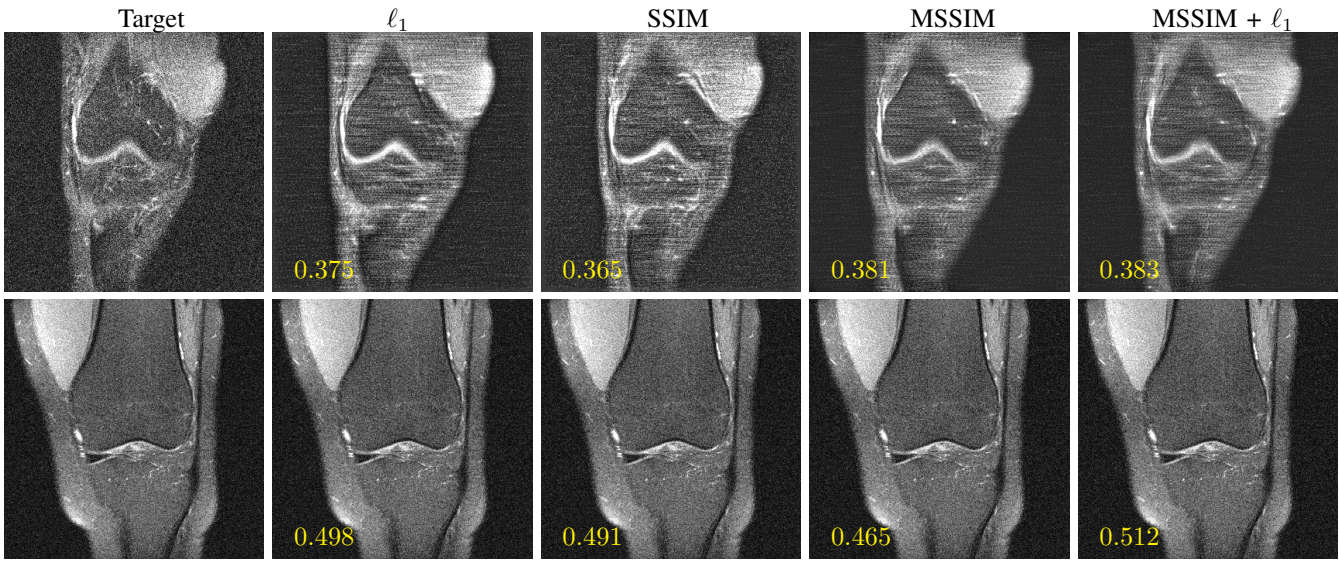

    \centering
    \include{images/qualitative_comparison/small_net}
    \caption{Two examples of single-coil 4x for the different loss functions. A small network is used to test several losses. SSIM values are shown in yellow.}
    \label{fig:loss_qualitative}
\end{figure*}

Table \ref{table:lossfuncs} reports the quantitative results for the different loss functions. The weighted linear combination of MSSIM and $\ell_1$ yielded the best results, where the $p$-values indicate that the improvement \markdown{achieved thanks to our modifications is highly consistent across all scans, despite the }small improvements on SSIM-values. Fig. \ref{fig:loss_qualitative} shows two example results for the different loss functions, confirming the favorable results for the model trained using a combination of MSSIM and $\ell_1$. Therefore, this loss function was selected for training the final model. For the remainder of the experiments, MSSIM is used as loss function. 

\begin{table}[!tb]
\let\center\empty
\let\endcenter\relax
\centering
\caption{
The effect of adopting a 2.5D approach on the 8x single-coil data using the small model. W denotes the loss weight applied to the neighboring slices. Stars denote one-way ANOVA statistical significance.}\label{table:multi_slice}
\resizebox{1.\width}{!}{\begin{tabular}{lllllll}
 \hline
 \multirow{2}{*}{Network} & \multicolumn{1}{c}{SSIM} & \multicolumn{1}{c}{NMSE} & \multicolumn{1}{c}{PSNR} & \multirow{2}{*}{$p$-value} \\
 & \multicolumn{1}{c}{$\mu\pm\sigma$} & \multicolumn{1}{c}{$\mu\pm\sigma$} & \multicolumn{1}{c}{$\mu\pm\sigma$} & \\
 \hline
\rowcolor{Gray}
2D    & {0.671$\pm$0.143$^*$} & {0.050$\pm$0.034} & {30.1$\pm$3.1} &$\ll$0.001\\
2.5D W0.1  & {0.549$\pm$0.128$^*$} & {0.089$\pm$0.034} & {26.8$\pm$2.1} &$\ll$0.001\\
\rowcolor{Gray}
2.5D W0.2  & {0.548$\pm$0.128$^*$} & {0.090$\pm$0.033} & {26.8$\pm$2.1} &$\ll$0.001\\

2.5D &\textbf{0.674$\pm$0.143} & \textbf{0.048$\pm$0.033} & \textbf{30.3$\pm$3.1} &\\
\hline
\end{tabular}

}
\end{table}

\subsection{Multi-slice Network}\label{sec:multi_slice}

The resolution of the images in the dataset is anisotropic with a voxel size of $0.5 \times 0.5 \times 3$ $mm^{3}$. Due to the correlation between adjacent slices with respect to anatomical structures in MRI images, we performed an experiment to assess whether inclusion of neighbouring slices into the reconstruction might improve the performance. We compared the 2D scheme using only the center slice with three alternative 2.5D schemes: i) the neighboring slices were used together with the center slice as input, but only the center slice was used in the loss function (network 2.5D); ii) and iii) the neighboring slices are also used in the loss, with different weights (0.1 vs 0.2 for the neighbors; 1.0 for the center slice). 
To compute the first and last slice, we pad the volume with replicas of the edge slices.
MSSIM was used for the loss function, 10 blocks, RMSprop as the optimizer, and 16 feature maps.

Table \ref{table:multi_slice} shows the results of this experiment, showing that the 2.5D schema very consistently improves over the 2D scheme, and that the loss should only be defined on the center slice. For the final model, this scheme was selected.

\subsection{Optimizer}\label{sec:optimizer}
We experimented with different optimizers including {RMSprop}, rectified Adam (RAdam) \cite{liu2019variance}, LookAhead \cite{zhang2019lookahead} and Ranger \cite{tong2019calibrating}. RAdam exploits a dynamic rectifier to adjust the adaptive momentum of Adam \cite{kingma2014adam}. LookAhead not only uses an adaptive learning rate and accelerated schemes but also iteratively updates two sets of weights, i.e. fast and slow weights. Ranger combines Radam and LookAhead optimizers into a single one. We used the 2D network with 10 blocks and 16 feature maps for each layer, and MSSIM the loss function.

Table \ref{table:optimizer_table} tabulates the results for the different optimizers. Since the best results were obtained for the RAdam optimizer, very consistently improving over the other optimizers, this was used for the final network. 

\begin{table}[!tb]
\let\endcenter\relax
\centering
\caption{The effect of the optimizer on performance, using the 8x single-coil validation data. Stars denote one-way ANOVA statistical significance. 
}\label{table:optimizer_table}
\resizebox{1.\width}{!}{\begin{tabular}{lllll}
\hline
\multirow{2}{*}{Optimizer} & \multicolumn{1}{c}{SSIM} & \multicolumn{1}{c}{NMSE} & \multicolumn{1}{c}{PSNR} & \multirow{2}{*}{$p$-value}\\ 
 & \multicolumn{1}{c}{$\mu\pm\sigma$} & \multicolumn{1}{c}{$\mu\pm\sigma$} & \multicolumn{1}{c}{$\mu\pm\sigma$} & \\
\hline
\rowcolor{Gray}
RMSprop   & {0.673$\pm$0.143$^*$} & {0.048$\pm$0.033} & {30.3$\pm$3.1} &$\ll$0.001\\
LookAhead & {0.668$\pm$0.140$^*$} & {0.050$\pm$0.032} & {30.0$\pm$2.9} &$\ll$0.001\\
\rowcolor{Gray}
Ranger    & {0.668$\pm$0.140$^*$} & {0.050$\pm$0.032} & {30.0$\pm$2.9} &$\ll$0.001\\
RAdam     & \textbf{{0.674}$\pm$0.141}& \textbf{{0.048$\pm$0.032}}& \textbf{{30.3$\pm$3.0}} &\\  \hline
\end{tabular}
}
\end{table}

\subsection{Adaptive-CS-Net vs ISTA-Net\texorpdfstring{$^{+}$}{+}}\label{sec:vs_ISTANET}

\begin{figure*}[!tb]
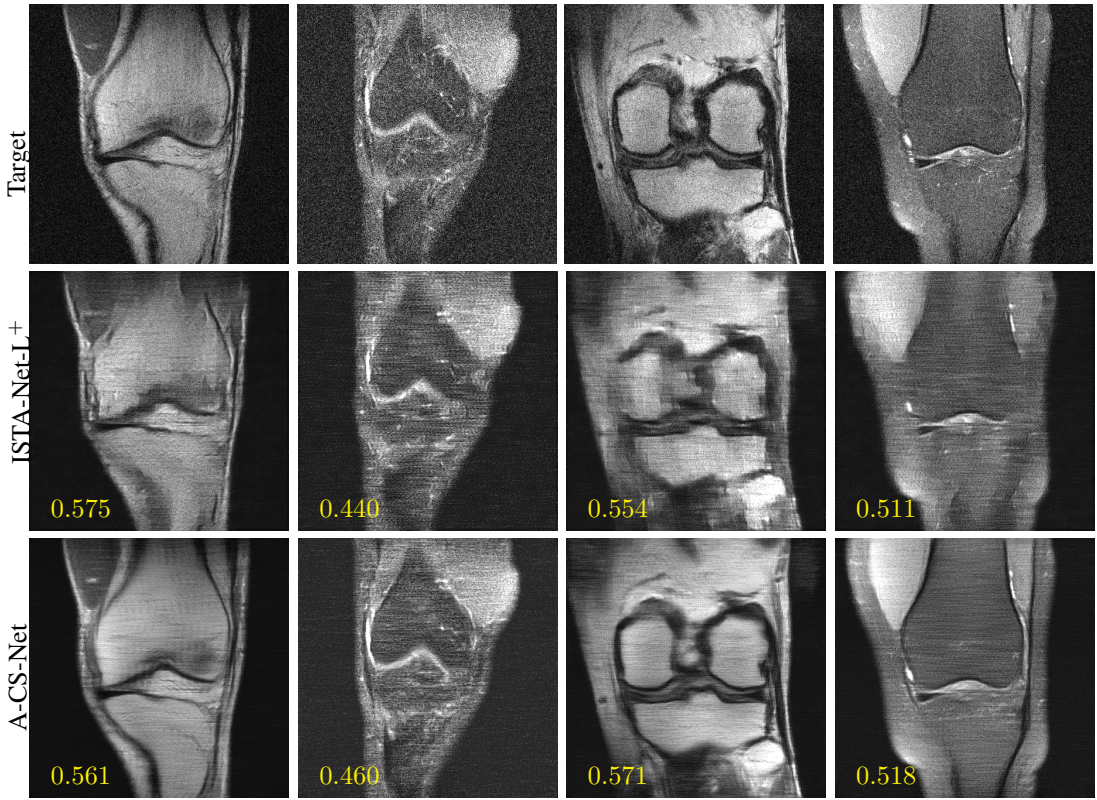

    \centering
    \include{images/ista_adaptive}
    \caption{Qualitative comparison of Adaptive-CS-Net vs ISTA-Net$^{+}$ on the single-coil 4x dataset. The SSIM values are shown in yellow.}
    \label{fig:ista_adaptive}
\end{figure*}

In this experiment, we compare the proposed model to ISTA-Net$^{+}$~\cite{zhang2018ista}. For this experiment, a 2D network with 10 blocks and 16 feature maps per layer was used, SSIM as loss function, and RAdam as the optimizer. Since ISTA-Net$^{+}$ uses a much smaller single scale architecture with much fewer network parameters, we added an experiment increasing the feature maps for ISTA-Net$^{+}$ such that the number of parameters was the same as for our architecture.
According to the results reported in Table \ref{table:istaplus}, the proposed model outperforms ISTA-Net$^{+}$ significantly. Figure \ref{fig:ista_adaptive} shows a qualitative comparison between ISTA-Net$^{+}$ and Adaptive-CS-Net on the single-coil 4x dataset. Although for the first image Adaptive-CS-Net reconstructed a better output in terms of the anatomical structure, the output of ISTA-Net-L$^+$ has a higher SSIM value. This implies that the radiological evaluation is a complementary step to judge the quality of the results.

\begin{table}[tb]
\let\center\empty
\let\endcenter\relax
\centering
\caption{Adaptive-CS-Net vs ISTA-Net$^{+}$ on the 8x single-coil dataset. ISTA-Net$^+$ has 0.75M trainable parameters, while ISTA-Net-L$^+$ and A-CS-Net have 2.12M trainable parameters. Stars denote one-way ANOVA statistical significance.
}\label{table:istaplus}
\resizebox{1.\width}{!}{\begin{tabular}{lllllll}
\hline
\multirow{2}{*}{Model} & \multicolumn{1}{c}{SSIM} & \multicolumn{1}{c}{NMSE} & \multicolumn{1}{c}{PSNR} & \multirow{2}{*}{$p$-value} \\ 
 & \multicolumn{1}{c}{$\mu\pm\sigma$} & \multicolumn{1}{c}{$\mu\pm\sigma$} & \multicolumn{1}{c}{$\mu\pm\sigma$} & \\
\hline
\rowcolor{Gray}
ISTA-Net$^{+}$   & {0.547$\pm$0.117$^*$} & {0.169$\pm$0.022} & {23.8$\pm$1.9} & $\ll$0.001\\
ISTA-Net-L$^{+}$  & {0.543$\pm$0.119$^*$} & {0.103$\pm$0.038} & {26.2$\pm$2.0} & $\ll$0.001\\
\rowcolor{Gray}
A-CS-Net  & \textbf{0.671$\pm$0.143} & \textbf{0.050$\pm$0.034} & \textbf{30.1$\pm$3.1} &\\ \hline
\end{tabular}
}
\end{table}

\subsection{Soft Priors}\label{sec:prior_knowledge}

To assess the contribution of the additional soft priors, we compared the full model against a version without known phase behaviour $\mathbf{e}_{\phi,i}$ and without background information $\mathbf{e}_{\mathrm{bg},i}$. Visually, we observed only small differences. To verify the differences in a realistic setting, we submitted the results to the public leaderboard of the fastMRI challenge. As shown in Table \ref{table:priors}, the network with all priors performed better in terms of the SSIM metric, although the results worsened in terms of NMSE and PSNR. Despite the fact that the improvement was minimal, we decided to adopt all priors for the final model to ensure our participation in the last challenge phase, since the selection was based on SSIM. 

\begin{table}[tb]
\let\center\empty
\let\endcenter\relax
\centering
\caption{The effect of adding priors to the final network on performance, using the multi-coil test data.
}\label{table:priors}
\resizebox{1.\width}{!}{\begin{tabular}{c|cccc}
\hline
Acceleration & prior & SSIM & NMSE & PSNR \\ \hline
\rowcolor{Gray}
\multirow{2}{*}{ \cellcolor{white}4x} & {$-$} & 0.772 & \textbf{0.025} & {30.98} \\ 
                    & {$+$} & \textbf{0.773} & {0.028} & \textbf{33.49} \\
\hline
\rowcolor{Gray}
\multirow{2}{*}{ \cellcolor{white}8x} & {$-$} & 0.674 & \textbf{0.038} & \textbf{30.90} \\
                    & {$+$} & \textbf{0.675} & 0.044 & 30.27 \\
\hline  
\end{tabular}
}
\end{table}

\section{Adaptive-CS-NET: Submitted Model}\label{sec:final_net}
In this section, we describe the configuration of the submitted model~\cite{pezzotti2019adaptive} and analyze the resulting reconstructions. The final performance is evaluated with the quantitative metrics on the test and challenge datasets, and by presenting the radiological scores for the challenge dataset as performed by the fastMRI challenge organizers. 

\begin{figure*}[!tb]
\begin{center}
\includegraphics[width=1\textwidth]{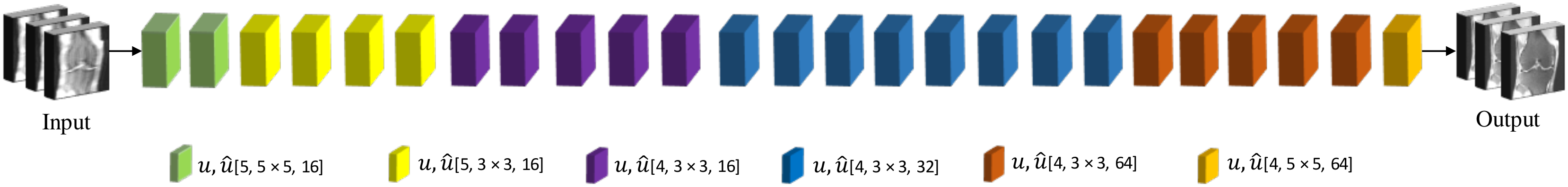}
\caption{Final network, each block has the same structure as shown in Fig. \ref{fig:our_solution_1} and is defined by $\mathcal{U}, \mathcal{\hat{U}}$[number of scales, kernel size, number of feature maps in the first scale]. For all layers Leaky ReLU was used as the activation function. }
\label{fig:net}
\end{center}
\end{figure*}

Following our model optimization study, the configuration of the final model was determined as follows. The linear combination of MSSIM and $\mathrm \ell_1$ \eqref{eq:loss_mssim_l1} was chosen as the loss function. The 2.5D scheme was chosen with two neighboring slices, with the loss applied only on the central slice. For training the model, the RAdam optimizer was deployed. Fig. \ref{fig:net} shows the structure of the final network. Each block is determined by three parameters for the denoiser: 1) the number of scales for the denoiser $\mathcal{U}, \mathcal{\hat{U}}$, 2) the kernel size used in the convolutions and, 3) the number of feature maps in the first convolutional layer, which is then doubled at each scale. According to the experiments presented in Fig.~\ref{fig:phaseno}, the number of reconstruction blocks greatly affects the reconstruction performance, empirically observing that performance still improves when 15 blocks are used. The available GPU memory is a limiting factor when designing a deep neural network. To allow for a large number of blocks, we chose a different design in each block, mixing a less powerful design (16 filters) with more powerful ones (64 filters). By adopting this strategy, our final design contained 25 reconstruction blocks \markdown{and has 33M parameters. }


\begin{table}[tb!]
\let\center\empty
\let\endcenter\relax
\centering
\caption{\label{table:final_metrics}Results for the final model for single- and multi-coil data on the test and challenge dataset.}
\resizebox{1.\width}{!}{
\begin{tabular}{l|l|l|l|l|l|l}
\hline
Dataset & Coil&& Detail & SSIM & NMSE & PSNR \\
\hline
\multirow{12}{*}{Test} & \multirow{6}{*}{multi} & \multirow{3}{*}{4x} & ALL & 0.928 & 0.005 &  39.9  \\
&& & PD   & 0.961 & 0.002 & 41.7 \\ 
&& & PDFS &0.891 & 0.009 & 37.9  \\ \cline{3-7}
&& \multirow{3}{*}{8x} & ALL & 0.888 & 0.009 & 36.8 \\ 
&& & PD   & 0.937 & 0.005 & 38.5 \\ 
&& & PDFS & 0.843 & 0.013 & 35.3 \\ 
\cline{2-7}
& \multirow{6}{*}{single} & \multirow{3}{*}{4x} & ALL & 0.777 &0.027 & 33.7 \\
&& & PD   & 0.877 & 0.010 &  36.9\\
&& & PDFS & 0.685 & 0.043 &  30.7  \\ \cline{3-7}
&& \multirow{3}{*}{8x} & ALL & 0.680 & 0.042 & 30.5  \\ 
&& & PD   & 0.777 & 0.019 & 32.4 \\ 
&& & PDFS & 0.575 & 0.067 & 28.5 \\ 
\hline
\multirow{3}{*}{Challenge} & \multirow{2}{*}{multi} & \multirow{1}{*}{4x} & ALL & 0.927 & 0.005 & 39.9  \\ 
\cline{3-7}
&& \multirow{1}{*}{8x} & ALL & 0.901 & 0.009 & 37.4 \\ 
\cline{2-7}
& \multirow{1}{*}{single} & \multirow{1}{*}{4x} &ALL & 0.751 & 0.030 & 32.7 \\ 
\hline  
\end{tabular}

}
\end{table}

\begin{figure*}[!tb]
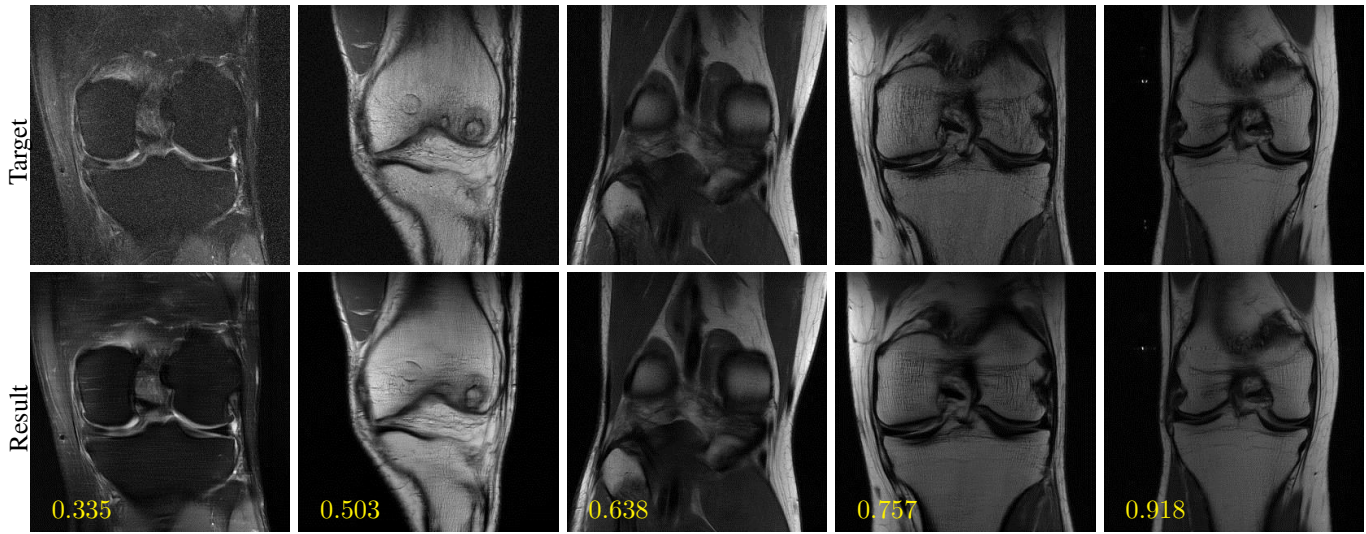

    \centering
    \include{images/qualitative_comparison/big_net_validationt}
    \caption{Example results of the final model for the multi-coil track accelerated by 8x on the validation dataset. Top row depicts the target image, bottom row the reconstructed images with the SSIM value in yellow. }
    \label{fig:big_net_validationt}
\end{figure*}

\begin{figure*}[!tb]
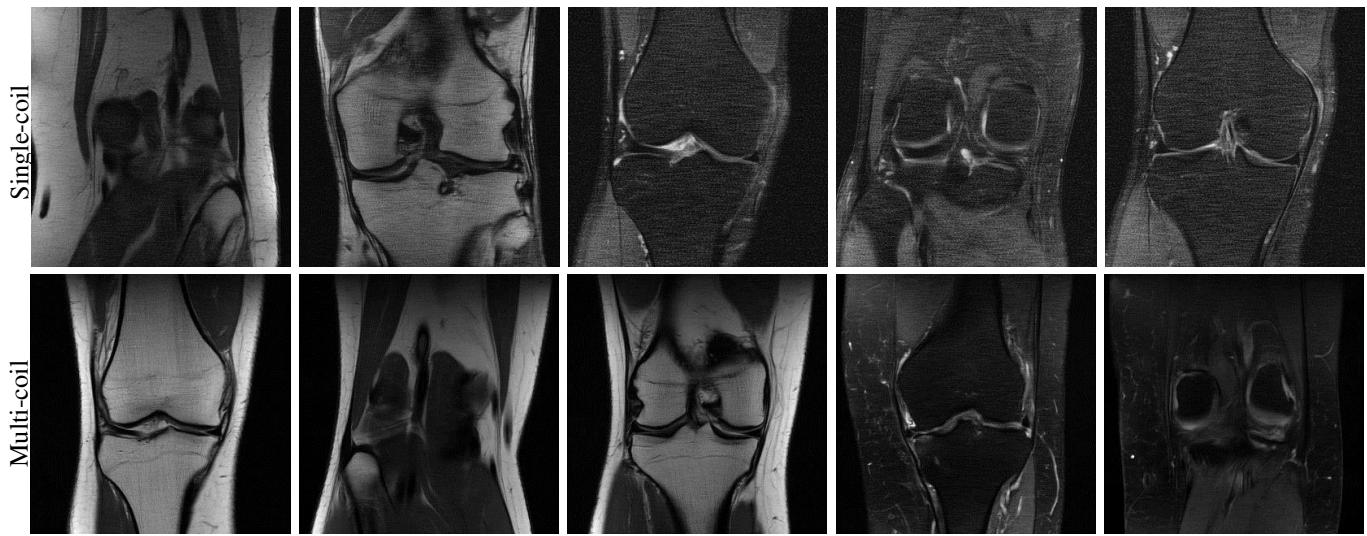

    \centering
    \include{images/qualitative_comparison/big_net_challenge}
    \caption{Example results of the final network from the test and challenge datasets, for which no ground truth reconstructions are available. }
    \label{fig:big_net_challenge}
\end{figure*}

Fig. \ref{fig:big_net_validationt} shows example results of the final network for the multi-coil track from the validation dataset. Fig. \ref{fig:big_net_challenge} shows examples from the test and challenge datasets.
Table \ref{table:final_metrics} shows the SSIM, NMSE, and PSNR values for the test and challenge set (as described in Section \ref{sec:dataset}), for the images with and without fat suppression and both combined, for both single- and multi-coil MRI scans. For the radiological evaluation, our method scored 2.285, 1.286, and 2.714 for multi-coil 4x, multi-coil 8x, and single-coil 4x, respectively (the closer to 1, the better). \markdown{The average runtimes for the model are 518 and 327 milliseconds for the multi-coil and the single-coil data, respectively.} More details on the results for the challenge were presented in \cite{knoll2020advancing}. 

\section{Discussion}

In this paper we propose a general method, named Adaptive-CS-Net, for reconstructing undersampled MRI data, combining ideas from compressed sensing theory with ideas from MR physics and deep learning. The method was developed in the context of the 2019 fastMRI challenge, which focused on accelerating knee MR imaging. 
The proposed network is an unrolled iterative learning-based reconstruction scheme, in which a large number of reconstruction blocks refine the MR image by denoising the signal in a learned and multi-scale fashion. 
Moreover, we added neighboring slices as input to the sparsifying transform, as well as a number of soft priors that encode MRI domain knowledge.

The main driver of the performance of our network is the multi-scale architecture, as demonstrated in a direct comparison with ISTA-Net$^+$ that is corrected by the number of trainable parameters.
According to the experimental results on the number of  blocks for 4x and 8x accelerations of both single- and multi-coil data, we showed that the number of blocks has a large impact on model performance. Therefore, it was decided to use the maximum number of blocks that we could fit into the GPU memory, where we adopted different model designs for the different blocks to save memory. 
It might be expected that beyond a certain number of blocks, overfitting of the data might occur. However, signs of overfitting were not observed during training and the final number of blocks was only marginally larger than tested in the optimization experiments. Whether further increase in the number of blocks could \markdown{result} in even better performance could be the topic of further experiments. This would, however, need better hardware, as the current design is memory- and time-bound during training. With the current configuration, final model training took approximately 7 days on two V100 GPUs.

We experimented with a large variety of loss functions. Results showed that the linear summation of MSSIM and $\ell_1$ performed best. Figure~\ref{fig:big_net_validationt} showed that poor SNR data yield very low SSIM scores. Surprisingly, within high SNR data, a large variance of SSIM scores is also found. This highlights the fact that further research is required in order to develop better quality metrics. Moreover, we defined a 2.5D scheme to train the network in which three adjacent slices were reconstructed while the loss function was calculated only for the central slice. The proposed scheme outperformed the 2D network as well as 2.5D networks in which the loss was calculated over all slices. By incorporating the neighbouring slices, the network can exploit existing correlations into the reconstruction of the target slice, which is our main target as defined by our loss. It can be expected that for MRI acquisition with less asymmetric voxel sizes, the inclusion of information of neighbouring slices would become more important. \markdown{However, weighing the loss of the neighbouring slices resulted in less optimal results since it forces the network to solve a more difficult problem: the network has to reconstruct multiple slices instead of a single one. This reduces the effective network capacity per slice, leading to a degradation of the reconstruction performance.}
We tested different optimizers, where the newly introduced RAdam outperformed the others and we used it for training the final network. 
We also incorporated prior knowledge, including data consistency, known phase behaviour and background discrimination to support the network in the reconstruction process. We observed that these priors provided only limited extra performance to the network, resulting in visually similar images and minimal difference in the metrics.

We can conclude that the Adaptive-CS-Net is sufficiently powerful to learn directly from the data how to reconstruct the undersampled k-space, being the multi-scale structure and the use of many reconstruction blocks the main driver of our performance.
As a future work, we want to better understand how much the network is relying on the priors by adopting interpretable AI techniques such as differentiable image parameterizations for feature visualization~\cite{mordvintsev2018differentiable}. Stronger use of the priors via the loss function is an additional option. 


As mentioned before, the radiologist scores were based on the visual quality of the reconstructed images and not on diagnostic interchangeability. Therefore, designing a network based on the diagnosis can be considered a point for further research. We furthermore observed that optimizing for SSIM was needed for reaching the final stage of the challenge, but is not necessarily an ideal representative of radiological image quality. This observation was very recently confirmed in a comparative study by others \cite{Mason2020}. The proposed method outperforms the benchmark networks, including U-net \cite{ronneberger2015u}, cascade net \cite{schlemper2017deep}, KIKI-net \cite{eo2018kiki}, and PD-net \cite{adler2018learned}, on the single-coil track as reported in \cite{ramzi2020benchmarking}. It outperforms as well the i-RIM model~\cite{putzky2019invert} on the  Multi-coil track but not the single coil track~\cite{knoll2020advancing}.

\section{Conclusion}

In this paper we propose an adaptive intelligence algorithm called Adaptive-CS-Net, which was developed in the context of the 2019 fastMRI challenge. In the two clinically relevant tracks of the challenge, using multi-coil MRI acquisitions, the proposed method was leading, while on a simulated single-coil track the method ranked 3rd.
\section{Acknowledgements}
Facebook AI Research and NYU Langone Health are acknowledged for organizing the 2019 fastMRI challenge.
\bibliographystyle{ieeetr}
\bibliography{references}

\begin{thebibliography}{10}

\bibitem{zaitsev2015motion}
M.~Zaitsev, J.~Maclaren, and M.~Herbst, ``Motion artifacts in {MRI}: A complex
  problem with many partial solutions,'' {\em Journal of Magnetic Resonance
  Imaging}, vol.~42, no.~4, pp.~887--901, 2015.

\bibitem{cohen1991ultra}
M.~S. Cohen and R.~M. Weisskoff, ``Ultra-fast imaging,'' {\em Magnetic
  Resonance Imaging}, vol.~9, no.~1, pp.~1--37, 1991.

\bibitem{donoho2006compressed}
D.~L. Donoho, ``Compressed sensing,'' {\em IEEE Transactions on Information
  Theory}, vol.~52, no.~4, pp.~1289--1306, 2006.

\bibitem{lustig2007sparse}
M.~Lustig, D.~Donoho, and J.~M. Pauly, ``Sparse {MRI}: The application of
  compressed sensing for rapid {MR} imaging,'' {\em Magnetic Resonance in
  Medicine}, vol.~58, no.~6, pp.~1182--1195, 2007.

\bibitem{candes2011compressed}
E.~J. Candes, Y.~C. Eldar, D.~Needell, and P.~Randall, ``Compressed sensing
  with coherent and redundant dictionaries,'' {\em Applied and Computational
  Harmonic Analysis}, vol.~31, no.~1, pp.~59--73, 2011.

\bibitem{liang1992constrained}
Z.-P. Liang, F.~Boada, R.~Constable, E.~Haacke, P.~Lauterbur, and M.~Smith,
  ``Constrained reconstruction methods in {MR} imaging,'' {\em Reviews in
  Magnetic Resonance in Medicine}, vol.~4, no.~2, pp.~67--185, 1992.

\bibitem{pruessmann1999sense}
K.~P. Pruessmann, M.~Weiger, M.~B. Scheidegger, and P.~Boesiger, ``{SENSE}:
  sensitivity encoding for fast {MRI},'' {\em Magnetic Resonance in Medicine},
  vol.~42, no.~5, pp.~952--962, 1999.

\bibitem{liang2019deep}
D.~Liang, J.~Cheng, Z.~Ke, and L.~Ying, ``Deep {MRI} reconstruction: Unrolled
  optimization algorithms meet neural networks,'' {\em arXiv preprint
  arXiv:1907.11711}, 2019.

\bibitem{quan2018compressed}
T.~M. Quan, T.~Nguyen-Duc, and W.-K. Jeong, ``Compressed sensing {MRI}
  reconstruction using a generative adversarial network with a cyclic loss,''
  {\em IEEE Transactions on Medical Imaging}, vol.~37, pp.~1488--1497, 2018.

\bibitem{mardani2018deep}
M.~Mardani, E.~Gong, J.~Y. Cheng, S.~S. Vasanawala, G.~Zaharchuk, L.~Xing, and
  et~al, ``Deep generative adversarial neural networks for compressive sensing
  {MRI},'' {\em IEEE Transactions on Medical Imaging}, vol.~38, no.~1,
  pp.~167--179, 2018.

\bibitem{guo2020deep}
Y.~Guo, C.~Wang, H.~Zhang, and G.~Yang, ``Deep attentive wasserstein generative
  adversarial networks for mri reconstruction with recurrent
  context-awareness,'' {\em arXiv preprint arXiv:2006.12915}, 2020.

\bibitem{schlemper2018stochastic}
J.~Schlemper, G.~Yang, P.~Ferreira, A.~Scott, L.-A. McGill, Z.~Khalique,
  M.~Gorodezky, M.~Roehl, J.~Keegan, D.~Pennell, {\em et~al.}, ``Stochastic
  deep compressive sensing for the reconstruction of diffusion tensor cardiac
  mri,'' in {\em International conference on medical image computing and
  computer-assisted intervention}, pp.~295--303, Springer, 2018.

\bibitem{yang2017dagan}
G.~Yang, S.~Yu, H.~Dong, G.~Slabaugh, P.~L. Dragotti, X.~Ye, F.~Liu,
  S.~Arridge, J.~Keegan, Y.~Guo, {\em et~al.}, ``Dagan: Deep de-aliasing
  generative adversarial networks for fast compressed sensing mri
  reconstruction,'' {\em IEEE transactions on medical imaging}, vol.~37, no.~6,
  pp.~1310--1321, 2017.

\bibitem{yu2017deep}
S.~Yu, H.~Dong, G.~Yang, G.~Slabaugh, P.~L. Dragotti, X.~Ye, F.~Liu,
  S.~Arridge, J.~Keegan, D.~Firmin, {\em et~al.}, ``Deep de-aliasing for fast
  compressive sensing mri,'' {\em arXiv preprint arXiv:1705.07137}, 2017.

\bibitem{putzky2019rim}
P.~Putzky, D.~Karkalousos, J.~Teuwen, N.~Miriakov, B.~Bakker, M.~Caan, and
  M.~Welling, ``{i-RIM} applied to the {fastMRI} challenge,'' {\em arXiv
  preprint arXiv:1910.08952}, 2019.

\bibitem{putzky2019invert}
P.~Putzky and M.~Welling, ``Invert to learn to invert,'' in {\em Advances in
  Neural Information Processing Systems}, pp.~446--456, 2019.

\bibitem{zhu2018image}
B.~Zhu, J.~Z. Liu, S.~F. Cauley, B.~R. Rosen, and M.~S. ~, ``Image
  reconstruction by domain-transform manifold learning,'' {\em Nature},
  vol.~555, no.~7697, pp.~487--492, 2018.

\bibitem{lee2018deep}
D.~Lee, J.~Yoo, S.~Tak, and J.~C. Ye, ``Deep residual learning for accelerated
  {MRI} using magnitude and phase networks,'' {\em IEEE Transactions on
  Biomedical Engineering}, vol.~65, no.~9, pp.~1985--1995, 2018.

\bibitem{wang2018image}
G.~Wang, J.~C. Ye, K.~Mueller, and J.~A. Fessler, ``Image reconstruction is a
  new frontier of machine learning,'' {\em IEEE Transactions on Medical
  Imaging}, vol.~37, no.~6, pp.~1289--1296, 2018.

\bibitem{hammernik2018learning}
K.~Hammernik, T.~Klatzer, E.~Kobler, M.~P. Recht, D.~K. Sodickson, T.~Pock, and
  F.~Knoll, ``Learning a variational network for reconstruction of accelerated
  {MRI} data,'' {\em Magnetic Resonance in Medicine}, vol.~79, no.~6,
  pp.~3055--3071, 2018.

\bibitem{sun2016deep}
J.~Sun, H.~Li, {\em et~al.}, ``Deep {ADMM-Net} for compressive sensing {MRI},''
  in {\em Advances in Neural Information Processing Systems}, pp.~10--18, 2016.

\bibitem{yang2010fast}
J.~Yang, Y.~Zhang, and W.~Yin, ``A fast alternating direction method for
  {TVL1-L2} signal reconstruction from partial {Fourier} data,'' {\em IEEE
  Journal of Selected Topics in Signal Processing}, vol.~4, no.~2,
  pp.~288--297, 2010.

\bibitem{aggarwal2018modl}
H.~K. Aggarwal, M.~P. Mani, and M.~Jacob, ``{MoDL}: Model-based deep learning
  architecture for inverse problems,'' {\em IEEE Transactions on Medical
  Imaging}, vol.~38, no.~2, pp.~394--405, 2018.

\bibitem{ramzi2020benchmarking}
Z.~Ramzi, P.~Ciuciu, and J.-L. Starck, ``Benchmarking deep nets {MRI}
  reconstruction models on the {fastMRI} publicly available dataset,'' in {\em
  2020 IEEE 17th International Symposium on Biomedical Imaging (ISBI)},
  pp.~1441--1445, IEEE, 2020.

\bibitem{zbontar2018fastMRI}
J.~Zbontar, F.~Knoll, A.~Sriram, M.~J. Muckley, M.~Bruno, A.~Defazio,
  M.~Parente, K.~J. Geras, J.~Katsnelson, H.~Chandarana, {\em et~al.},
  ``{fastMRI}: An open dataset and benchmarks for accelerated mri,'' {\em arXiv
  preprint arXiv:1811.08839}, 2018.

\bibitem{ronneberger2015u}
O.~Ronneberger, P.~Fischer, and T.~Brox, ``U-net: Convolutional networks for
  biomedical image segmentation,'' in {\em International Conference on Medical
  Image Computing and Computer Assisted Intervention}, pp.~234--241, Springer,
  2015.

\bibitem{schlemper2017deep}
J.~Schlemper, J.~Caballero, J.~V. Hajnal, A.~Price, and D.~Rueckert, ``A deep
  cascade of convolutional neural networks for {MR} image reconstruction,'' in
  {\em International Conference on Information Processing in Medical Imaging},
  pp.~647--658, Springer, 2017.

\bibitem{eo2018kiki}
T.~Eo, Y.~Jun, T.~Kim, J.~Jang, H.-J. Lee, and D.~Hwang, ``{KIKI-net}:
  cross-domain convolutional neural networks for reconstructing undersampled
  magnetic resonance images,'' {\em Magnetic Resonance in Medicine}, vol.~80,
  no.~5, pp.~2188--2201, 2018.

\bibitem{adler2018learned}
J.~Adler and O.~{\"O}ktem, ``Learned primal-dual reconstruction,'' {\em IEEE
  Transactions on Medical Imaging}, vol.~37, no.~6, pp.~1322--1332, 2018.

\bibitem{caballero2014dictionary}
J.~Caballero, A.~N. Price, D.~Rueckert, and J.~V. Hajnal, ``Dictionary learning
  and time sparsity for dynamic {MR} data reconstruction,'' {\em IEEE
  Transactions on Medical Imaging}, vol.~33, no.~4, pp.~979--994, 2014.

\bibitem{chambolle2011first}
A.~Chambolle and T.~Pock, ``A first-order primal-dual algorithm for convex
  problems with applications to imaging,'' {\em Journal of Mathematical Imaging
  and Vision}, vol.~40, no.~1, pp.~120--145, 2011.

\bibitem{seitzer2018adversarial}
M.~Seitzer, G.~Yang, J.~Schlemper, O.~Oktay, T.~W{\"u}rfl, V.~Christlein,
  T.~Wong, R.~Mohiaddin, D.~Firmin, J.~Keegan, {\em et~al.}, ``Adversarial and
  perceptual refinement for compressed sensing mri reconstruction,'' in {\em
  International conference on medical image computing and computer-assisted
  intervention}, pp.~232--240, Springer, 2018.

\bibitem{zhang2018ista}
J.~Zhang and B.~Ghanem, ``{ISTA-Net}: Interpretable optimization-inspired deep
  network for image compressive sensing,'' in {\em Proceedings of the IEEE
  Conference on Computer Vision and Pattern Recognition}, pp.~1828--1837, 2018.

\bibitem{knoll2020advancing}
F.~Knoll, T.~Murrell, A.~Sriram, N.~Yakubova, J.~Zbontar, M.~Rabbat,
  A.~Defazio, M.~J. Muckley, D.~K. Sodickson, {\em et~al.}, ``Advancing machine
  learning for mr image reconstruction with an open competition: Overview of
  the 2019 fastmri challenge,'' {\em Magnetic Resonance in Medicine}, 2020.

\bibitem{knoll2020fastmri}
F.~Knoll, J.~Zbontar, A.~Sriram, M.~J. Muckley, M.~Bruno, A.~Defazio, and
  et~al, ``{fastMRI}: A publicly available raw k-space and {DICOM} dataset of
  knee images for accelerated {MR} image reconstruction using machine
  learning,'' {\em Radiology: Artificial Intelligence}, vol.~2, no.~1, 2020.

\bibitem{wang2004image}
Z.~Wang, A.~C. Bovik, H.~R. Sheikh, and E.~P. Simoncelli, ``Image quality
  assessment: from error visibility to structural similarity,'' {\em IEEE
  Transactions on Image Processing}, vol.~13, no.~4, pp.~600--612, 2004.

\bibitem{beck2009fast}
A.~Beck and M.~Teboulle, ``A fast iterative shrinkage-thresholding algorithm
  for linear inverse problems,'' {\em {SIAM} Journal on Imaging Sciences},
  vol.~2, no.~1, pp.~183--202, 2009.

\bibitem{donoho2009message}
D.~L. Donoho, A.~Maleki, and A.~Montanari, ``Message-passing algorithms for
  compressed sensing,'' {\em Proceedings of the National Academy of Sciences},
  vol.~106, no.~45, pp.~18914--18919, 2009.

\bibitem{araujo2019computing}
A.~Araujo, W.~Norris, and J.~Sim, ``Computing receptive fields of convolutional
  neural networks,'' {\em Distill}, vol.~4, no.~11, p.~e21, 2019.

\bibitem{paszke2017automatic}
A.~Paszke, S.~Gross, S.~Chintala, G.~Chanan, E.~Yang, Z.~DeVito, and et~al,
  ``Automatic differentiation in {PyTorch},'' in {\em Proceedings of Neural
  Information Processing Systems}, 2017.

\bibitem{johnson2016perceptual}
J.~Johnson, A.~Alahi, and L.~Fei-Fei, ``Perceptual losses for real-time style
  transfer and super-resolution,'' in {\em European Conference on Computer
  Vision}, pp.~694--711, Springer, 2016.

\bibitem{huber1992robust}
P.~J. Huber, ``Robust estimation of a location parameter,'' in {\em
  Breakthroughs in statistics}, pp.~492--518, Springer, 1992.

\bibitem{wang2003multiscale}
Z.~Wang, E.~P. Simoncelli, and A.~C. Bovik, ``Multiscale structural similarity
  for image quality assessment,'' in {\em The Thrity-Seventh Asilomar
  Conference on Signals, Systems \& Computers, 2003}, vol.~2, pp.~1398--1402,
  IEEE, 2003.

\bibitem{zhao2016loss}
H.~Zhao, O.~Gallo, I.~Frosio, and J.~Kautz, ``Loss functions for image
  restoration with neural networks,'' {\em IEEE Transactions on Computational
  Imaging}, vol.~3, no.~1, pp.~47--57, 2016.

\bibitem{zhao2015loss}
H.~Zhao, O.~Gallo, I.~Frosio, and J.~Kautz, ``Loss functions for neural
  networks for image processing,'' {\em arXiv preprint arXiv:1511.08861}, 2015.

\bibitem{liu2019variance}
L.~Liu, H.~Jiang, P.~He, W.~Chen, X.~Liu, J.~Gao, and et~al, ``On the variance
  of the adaptive learning rate and beyond,'' {\em ICLR}, 2020.

\bibitem{zhang2019lookahead}
M.~Zhang, J.~Lucas, J.~Ba, and G.~E. Hinton, ``Lookahead optimizer: k steps
  forward, 1 step back,'' in {\em Advances in Neural Information Processing
  Systems}, pp.~9593--9604, 2019.

\bibitem{tong2019calibrating}
Q.~Tong, G.~Liang, and J.~Bi, ``Calibrating the learning rate for adaptive
  gradient methods to improve generalization performance,'' {\em arXiv preprint
  arXiv:1908.00700}, 2019.

\bibitem{kingma2014adam}
D.~P. Kingma and J.~Ba, ``Adam: A method for stochastic optimization,'' {\em
  International Conference for Learning Representations}, 2015.

\bibitem{pezzotti2019adaptive}
N.~Pezzotti, E.~de~Weerdt, S.~Yousefi, M.~S. Elmahdy, J.~van Gemert,
  C.~Sch{\"u}lke, and et~al, ``{Adaptive-CS-Net}: {fastMRI} with adaptive
  intelligence,'' {\em arXiv preprint arXiv:1912.12259}, 2019.

\bibitem{mordvintsev2018differentiable}
A.~Mordvintsev, N.~Pezzotti, L.~Schubert, and C.~Olah, ``Differentiable image
  parameterizations,'' {\em Distill}, vol.~3, no.~7, p.~e12, 2018.

\bibitem{Mason2020}
A.~{Mason}, J.~{Rioux}, S.~E. {Clarke}, A.~{Costa}, M.~{Schmidt}, V.~{Keough},
  T.~{Huynh}, and S.~{Beyea}, ``Comparison of objective image quality metrics
  to expert radiologists’ scoring of diagnostic quality of {MR} images,''
  {\em IEEE Transactions on Medical Imaging}, vol.~39, no.~4, pp.~1064--1072,
  2020.

\end{thebibliography}
\clearpage

\end{document}